\documentclass[11pt]{article}
\usepackage{times, subfigure}
\usepackage{harvard}
\usepackage{url}
\usepackage{epsfig,setspace,rotating,fullpage}
\def\eqx"#1"{{\label{#1}}}
\def\eqn"#1"{{\ref{#1}}}

\usepackage{color}
\definecolor{gray}{rgb}{0.5,0.5,0.5}

\title{
I hear, I forget.  I do, I understand: \\ 
a modified Moore-method mathematical statistics course}
\author{Nicholas J. Horton\thanks{
Address for correspondence: Dept of Mathematics, Seeley Mudd, Amherst College,
Amherst, MA  01002-5000.  Phone: 413-542-5655, email:
nhorton@amherst.edu}
\\
\footnotesize 
Department of Mathematics\\
\footnotesize 
Amherst College, Amherst, MA
\normalsize
}
\begin{document}
\maketitle
\newpage
\begin{center}
{\large
I hear, I forget.  I do, I understand: \\ 
a modified Moore-method mathematical statistics course}
\end{center}

\subsection*{Abstract}

Moore introduced a method for graduate mathematics instruction that
consisted primarily of individual student work on 
challenging proofs \cite{jone:1977}.  \citeasnoun{cohe:1992} described an adaptation with less 
explicit competition suitable for
undergraduate students at a liberal arts college.
This paper details an adaptation of this modified Moore-method to teach mathematical
statistics, and describes ways that such an approach helps engage students and
foster the teaching of statistics.

Groups of students worked a set of 3 difficult problems (some theoretical, some applied)
every two weeks.
Class time was devoted to coaching sessions with the
instructor, group meeting time, and class presentations.  R was
used to estimate solutions empirically where analytic results were
intractable, as well as to provide an environment to undertake simulation
studies with the aim of deepening understanding and complementing analytic solutions.
Each group presented comprehensive solutions 
to complement oral presentations.
Development of parallel techniques for empirical and analytic problem solving was
an explicit goal of the course, which also
attempted to communicate ways that
statistics can be used to tackle interesting problems.
The group problem solving component and use of technology
allowed students to attempt much more challenging questions than
they could otherwise solve.

Keywords: 
capstone course,
empirical problem solving, 
intermediate statistics, 
R software,
RStudio integrated development environment,
reproducible analysis,
simulation studies,
statistical computing, 
statistical education

\newpage
\section{Introduction}

In this paper, an implementation of a mathematical statistics
course is described with the goal of developing
a combination
of analytic and empirical problem-solving skills through the solution
of challenging problems and complex
case studies. The course, offered at the 
Department of Mathematics and Statistics at Smith College in spring 2007 and
spring 2011, adapted the approach of
R.L Moore \cite{jone:1977}, using modifications suggested by 
\citeasnoun{cohe:1992}.  
A similar mathematical statistics course is described by 
\citeasnoun{mclo:2008}.

In the next subsection, 
recent developments in statistics education are described,
followed by an overview of the
modified Moore-Cohen method.
Section \ref{sec:structure} describes specific
details of the 
course, 
Section \ref{sec:problems} provides 
two example problems (with 
empirical as well as analytic solutions), 
Section \ref{sec:assessment} describes grading and assessment
while Section \ref{sec:discussion}
provides additional discussion and closing thoughts.

\subsection{Developments in statistical education}

Extensive curricular reforms in undergraduate statistics education
have transformed our programs and courses in recent decades \cite{cobb:1992,moor:cobb:1995,cobb:2011}.  The Guidelines for Assessment and Instruction
for Statistics Education (GAISE) report \cite{gaise}, which succinctly described 
these changes, 
recommended that introductory
statistics courses:
\begin{itemize}
\item Emphasize statistical literacy and develop statistical thinking,
\item Use real data,
\item Stress conceptual understanding rather than mere knowledge of procedures,
\item Foster active learning in the classroom,
\item Use technology for developing conceptual understanding and analyzing data, and
\item Use assessments to improve and evaluate student learning.
\end{itemize}

Other related efforts have attempted to broaden the types of questions
that statistics students grapple with \cite{brow:kass:2009,goul:2010},
increase the
use of case studies \cite{barr:tamb:1980,nola:spee:2000,nola:2003} and 
take advantage of sophisticated computing
technologies and environments such as R \cite{ihak:gent:1996} or
Matlab \cite{kapl:2003} to buttress understanding of statistical
concepts \cite{butt:nola:2001,nola:temp:2003,hort:brow:2004,froe:2008,nola:temp:2010,laza:2011}.

The mathematical statistics course has undergone many
transformations during this same period.  A lively panel in 2003 with the provocative title 
``Is the Math Stat Course obsolete?" \cite{obsolete} provided a glimpse into ways that this
intermediate level statistics course is adapting to a changing landscape.
One idea raised was that the math stat course (still a common entry point to the field for many students studying mathematics) should convey the excitement
of the discipline (``even if they don't go on [in statistics], we want them to 
leave thinking statistics is interesting"). Another was that modeling, computing and problem-solving are key components of such a course.

\citeasnoun{cobb:2011} 
provides a series of 
capsule summaries of innovations in the 
teaching of mathematical statistics, and discusses
key tensions that underlie our courses in terms
of what we want students to learn:
\begin{quote}
Surely the most common answer must be that we want our students to learn to analyze data,
and certainly I share that goal.  But for some students, particularly those with a strong
interest and ability in mathematics, I suggest a complementary goal, one that in my opinion
has not received enough explicit attention: We want these mathematically inclined students
to learn to solve methodological problems.  I call the two goals complementary because,
as I shall argue in detail, there are essential tensions between the goals of helping
students learn to analyze data and helping students learn to solve methodological problems.

For a ready example of the tension, consider the role of simple, artificial examples.  For teaching
data analysis, these ``toy'' examples are often and deservedly regarded with contempt.  But for 
developing an understanding of a methodological challenge, the ability to create a dialectical 
succession of toy examples and exploit their evolution is critical (p. 32).
\end{quote}

\subsection{Moore and Cohen methods}

Moore was noted \cite{halm:1985}
for quoting the Chinese proverb \emph{I hear, I forget.  I see,
I remember.  I do, I understand.} 
He provided classes with a list of definitions and theorems which they would 
subsequently prove individually and then share with the rest of the class.  
Competition was a key driving force in the course \cite{jone:1977}, with efforts to ensure that student background was as homogeneous as possible.  
The overall goal was to build student capacity to create structure from an axiomatic basis and communicate
this to others.
\citeasnoun{chri:2009} described possible
evaluations and assessment of Moore method mathematics courses.  

\citeasnoun{cohe:1992} modified Moore's approach using three guiding principles:
\begin{itemize}
\item Students understand better and remember longer what they discover
themselves than what is told to them,
\item People master an idea thoroughly when they teach it to someone else, and
\item Effective writing and clear thinking are inextricably linked (p. 474).
\end{itemize}
A fourth principle incorporated in this mathematical statistics course 
involved the use of 
R \cite{r-ref} and RStudio (an open source integrated development environment for R) to facilitate parallel empirical and analytic problem solving 
techniques.

Much of the class time 
is spent with students working as a group and individually
to solve sets of challenging problems, writing up solutions, and presenting them to the
class as a whole.  While each group tackled problems from each of the major
units of the course, group members would tend to learn their assigned problems in more detail
and rely on their classmates to convey understanding of the other problems.

The Moore and Cohen approaches deal more with pedagogy than with curriculum.
Moore used his method to teach proofs in topology.  Cohen used his
method for linear algebra.  Here we 
borrow Cohen's pedagogy for a course in mathematical 
statistics.  Students are not given theorems to prove as in Moore's
courses; instead they are given challenging problems to solve.

These problems are chosen according to four criteria.  The first
two are critical to the pedagogy: each problem should be easy to grasp, 
and each should be hard
enough that solving it poses a genuine challenge.  The first criterion
helps ensure that all students in a group can participate; the second
helps ensure that stronger students will not be able to cut off discussion
with a quick solution.

A third criterion is that the problems should have links to actual applications.
This is in the spirit of the GAISE
recommendations.  

Fourth, the problems should lend themselves to parallel and 
complementary pairs of solutions, one based on simulation and the other based 
on theory.  
The
parallel solutions constitute a recurring theme to the course, one
that is central to the curriculum.
This criterion is in some ways incidental to the pedagogy, 
although it helps ensure that students with different strengths and 
backgrounds can contribute actively to group work.  

\section{Details of the course}
\label{sec:structure}

For the sections (officially titled ``Seminar in Mathematical Statistics'') taught by 
the author in Spring 2007 and Spring 2011, the class met three times per week for 80 
minutes per session for thirteen weeks.  

While the only
required prerequisite for the class was probability,
most students in the course had also taken introductory statistics
and linear algebra. No specific knowledge of statistics was assumed.
At the beginning of the course, the students took the 
40 item multiple choice CAOS (Comprehensive
Assessment of Outcomes in a first statistics course) test
\cite{delm:garf:2006}.
While designed to assess student reasoning after a first course 
in statistics (and not a mathematical statistics class), the CAOS focuses 
on conceptual understanding of variability and uncertainty.
The average score for the mathematical statistics students on the CAOS 
pre-test was 67.2\% correct with a standard deviation of
13.6\% (values ranged from 43\% to 90\%).

The structure of the course included (almost) no lectures.  Instead, the
material was broken down into a number of problem sets.  These
questions were designed to be sufficiently difficult to provide 
a challenge to students, but still amenable 
(with some assistance)
to solution.

During the first offering, 
four groups of 3 students were 
created, with seven groups of 3 students for the 
second offering.  Each group would work a different set of problems for each 
problem set (with an occasional problem assigned to all groups).  Throughout the semester, these
groups were reshuffled twice, with no two individuals being in the same group
twice.  The re-balancing helped to address issues with groups that  consisted of only weak 
students (as well as to provide a release valve for problems with group
dynamics).  

Most class sessions consisted of a series of ``coaching'' sessions
several days after the initial presentation of the problems.  These coaching
sessions, described in detail in \citeasnoun{cohe:1992}, are critical
in helping to guide students towards the desired solution
without providing the answer.  
All students in a group attended a given coaching session, and discussed their
preliminary attempts at the problems.  In some cases they may have solved their
problems.
More commonly additional guidance was needed for them to make progress 
or to elaborate on their solutions.  Early on in the course, much of this coaching
involved support and scaffolding for the use of computing (to allow them to gradually
build their skills in terms of simulation and exploration).

Each student created a draft of their preliminary solution in preparation 
for a second
coaching session.  To ensure that all students were engaged and making good faith
efforts, these were reviewed by the instructor.
One per group was graded in detail, to provide
general feedback for all students.

The second coaching session was used to help answer any remaining questions and
assist with preparations of the solutions (``weekly papers'').  Other assistance
was provided outside of the regular class meeting times by email or during
office hours.

Before the final class session for a given set of problems,
each group created a single clear and
comprehensive solution, which was made available to the
class.  
Finally, each group gave a 15 minute oral presentation that reviewed their 
solution, with questions and answers as needed.

\subsection{Textbooks and topics}

The approach suggested by \citeasnoun{cohe:1992} to teach
analysis of linear algebra provides students 
several pages of axioms, definitions, theorems and problems.  This serves as the foundation
from which all of the remaining material is derived.
Because of the need for more extensive material to 
support student work in a range of mathematical statistical topics, 
the text by
\citeasnoun{rice:1995} was used for background reading as well as the source
of many of the problems.  In addition, several modules (including case studies
with advanced data analysis) were integrated from 
\citeasnoun{nola:spee:2000}.

The course began with a series of challenging probability problems,
covering selected topics and highlights from Chapters 1 through 5 of \citeasnoun{rice:1995}.
The next set of problems related to descriptive and graphical visualization 
(covering Chapter 10 of \citeasnoun{rice:1995} and the \emph{Maternal smoking and
infant health} module from \citeasnoun{nola:spee:2000}). Two sets of problems
were devoted to estimation  and the bootstrap (Chapter 8 of \citeasnoun{rice:1995} and the 
\emph{Patterns in DNA} and \emph{Who plays video games?} modules from 
\citeasnoun{nola:spee:2000}).  Testing hypotheses and assessing goodness of fit
(Chapter 9 of \citeasnoun{rice:1995}) comprised two sets of problems,
while Chapter 10 was used as a basis for a set on two sample comparisons.
The first time the course was offered, it closed with a set of problems entitled \emph{Bayesian inference:
a big idea}, based loosely on Chapter 15 of \citeasnoun{rice:1995} and
Section 2.5 of \citeasnoun{lavi:2007}, while the second offering closed with
precursors of informal inference and simulation studies of inference rules \cite{wild:2010}.

\subsection{Real data and mathematical statistics}

While the course did not focus on advanced analysis of
multivariate datasets, real data was regularly incorporated into
the course, primarily as a component of problems assigned to the
students throughout the semester.  The textbooks by Rice as
well as Nolan \& Speed are notable for the number and variety of
motivating examples provided throughout, including the exercises.
As an example, students might be asked to find the method of moments
estimator for $\theta$ for a Pareto distribution with known scale
parameter $x_0$, and compare this to the maximum likelihood estimator for $\theta$.
After finding the analytic results, and simulating to
compare the variance of the estimators, they would be asked to calculate
and interpret the sample statistic using data from an economic survey.
Another set of problems related to the analysis of cell probabilities
expected by genetic theories, through estimation of underlying parameters.
Students were assessed both on their ability to report in context
on the underlying applied statistical question, as well as on the
relevant statistical derivations or simulations that
they carried out.

As outlined by the GAISE guidelines,
use of real data is essential to the introductory course, and central
also to any statistics curriculum as a whole.  Nevertheless, for certain
individual courses that serve as elements of a larger statistics
curriculum, real data may be less essential.  There is an
inherent complementarity between analysis of data using existing methods
and the development of new methods \cite{cobb:2011}.
We 
need a curriculum that teaches students to engage, appreciate 
and enjoy both 
data analytic and methodological challenges.  
In a course such as the one described here, although
connections to real data are important, the balance is weighted towards 
problems of a more abstract sort.

\subsection{Technology}

This approach would not be feasible without the use
of computing technology to facilitate analysis and simulation.
R \cite{ihak:gent:1996} and RStudio 
(\url{http://rstudio.org})
provided a flexible and adaptable environment for
exploration \cite{hort:brow:2004,prui:2011}.

RStudio 
is an open-source integrated development environment that provides a consistent 
and powerful interface for R (an open source general purpose statistical package, \url{http://r-project.org})
that is easier to install, learn and run than standard R.
\LaTeX\ \cite{latex} within the Sweave \cite{leis:2002} system  was used as the formatting environment for the solutions, with
an annotated example distributed to all students during the initial
class meeting.  This included examples of tables, figures, cross referencing,
bibliography
and other useful attributes.  Submissions were made available to students as both
Sweave source and PDF files to allow students to borrow working code.
RStudio is particularly attractive because it simplifies the user interface and has tightly integrated support for 
Sweave (including a 
single button click to {\tt Compile PDF} from the source document).
In future offerings, the {\tt Markdown} system within the {\tt knitr} package \cite{knitr} will be used, as it provides simplified 
functionality and does not require knowledge of \LaTeX.

The course intentionally introduced students to concepts of \emph{reproducible analysis} \cite{gent:temp:2007}, where computation, code and results of an analysis are integrated.
Being able to re-run a set of simulations and regenerate a report with a single click is a powerful 
motivator for students used to error-prone processes of cutting and pasting output and figures.
Reproducible analysis systems are becoming standard in industry and academia, have the potential to help ensure
better statistical analysis, and should be incorporated in the statistics curriculum.

To help simplify the learning curve for these somewhat complex systems, 
a number
of examples and idioms were provided by the instructor,
to help build students' repertoire of useful techniques to attack problems.
Students were
encouraged to write their initial solutions using pseudo-code (an
informal description that 
could later be turned into working R code).  These were also posted to the course
management system to facilitate
re-use in other problems and settings.

\section{Example problems and solutions} \label{sec:problems}

To give a better sense of the course, we describe two problems that were completed by the students, along with model solutions and commentary
(additional
examples are found in the online supplement).  Each group would generally work 3 or 4 problems 
per assignment.  

These problems feature both empirical (simulations in R) and analytic (closed-form) solutions
by the groups.
They range from easier to more challenging, but illustrate
the approaches taken by students in three separate application areas.
The general level of difficulty is similar to
that of \citeasnoun{rice:1995}\footnote{Rice states on page xi that 
\emph{This book includes a fairly large number of problems, some of which will be quite 
difficult for students}. My students confirmed this assertion.}.  

\subsection{Pooled blood sera sampling} \label{sera}

It is known that 5\% of the members of a population have disease
X, which can be discovered by a blood test (that is assumed to 
perfectly identify both diseased and nondiseased populations).
Suppose that $N$ people
are to be tested, and the cost of the test is
non-trivial.  The testing can be done in two ways:
(A) Everyone can be tested separately; or (B) the blood samples of
$k$ people are pooled to be analyzed.  Assume that $N=nk$
with $n$ an integer.  If the test is negative, all the people tested are
healthy (that is, just this one test is needed).  If the test result is
positive, each of the $k$ people must be tested separately (that
is, a total of $k+1$ tests are needed for that group)\footnote{This assumes
that all of the tests are run at the same time.  Otherwise, if the pool tested positive and the first $k-1$ tests were negative, there would be no need to test the final member of the pool.}.

Questions: 
\begin{enumerate}
\item For fixed $k$ what is the
expected number of tests needed in (B)?  

\item Find the $k$ that will
minimize the expected number of tests in (B).  

\item Using the $k$ that minimizes
the number of tests,
on  average how many tests does (B) save in
comparison with (A)?  Be sure to check your answer using an empirical
simulation.
\end{enumerate}

\subsubsection{Empirical (simulation-based) solution}

We attempted to gain a better understanding of the problem by
simulation. First, we set $k=10$, $n=500$ and $P(\textrm{infected})=p=0.05$ (refer to Figure
\ref{fig:code1a} for code). Given these specific values for each of
the variables, we found the expected number of tests to be approximately
2501.9. We then used this value to help us check our analytic
solution.

Next, we tried different values of $k$ and $n$ such that $N$ 
(the number of people to be tested) equaled
5000. We did this to find the value of $k$ that
minimized the expected number of tests. Given that
$N=5000$, possible integer values for $k$ were 2, 4, 5, 8, and 10.
We found the expected number
of tests for each of these $k$ values, respectively were 2988.7,
2178.9, 2126.5, 2306.2, and 2501.9. For this example, the
minimum value is found when $k=5$.

\begin{figure}[htpb]

\begin{verbatim}
numsim = 1000
p = 0.05       # probability of infection
k = 10         # size of pool
n = 500        # number of pools
N = n*k        # total number of people

# for Approach (A), E[# tests] = N
# for Approach (B), find E[# tests] if pooled?
res = numeric(numsim)    # stash results
for (i in 1:numsim) {
  disease = rbinom(N, size=1, prob=p)
  poolmat = matrix(disease, nrow=k, ncol=n)
  Y = apply(poolmat, 2, sum)>=1
  res[i] = n+sum(Y)*k
}
mean(res)
sd(res)
curve(N/x+x*(N/x)*(1-(1-p)^x), xlab="number in pool (k)", 
  ylab="expected number", from=2, to=10)
\end{verbatim}
\caption{R code to generate empirical estimates using Approach (B)}%
\label{fig:code1a}
\end{figure}

\begin{figure}
\begin{center}
\includegraphics[height=3.8in,width=4.8in]{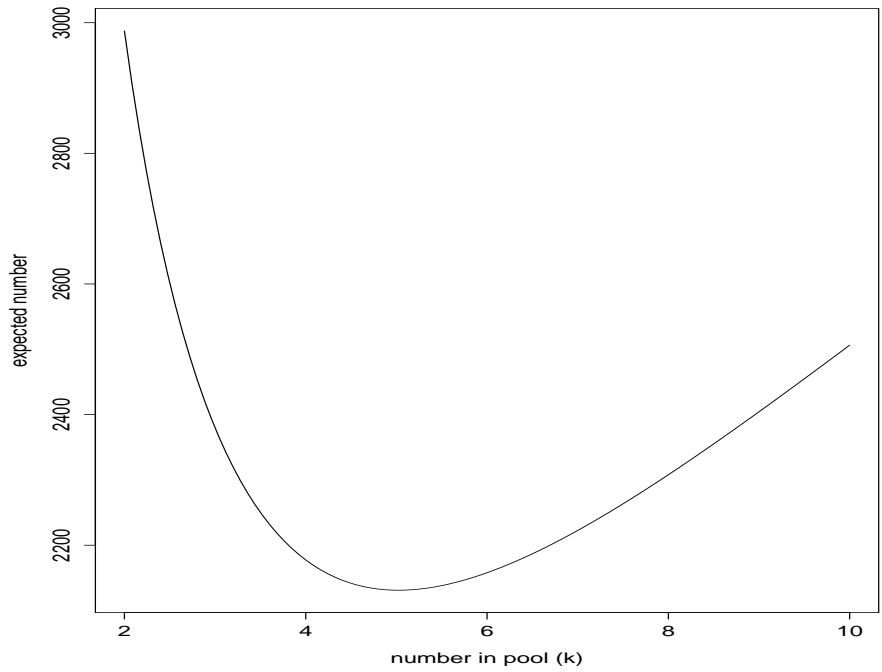}
\caption{Display of expected number of blood tests required as a function of pool-size ($k$), with $N=5000, p=0.05$}
\label{fig:plot1b}
\end{center}
\end{figure}

\subsubsection{Analytic (closed-form) solution}

Approach (A):
the expected number of tests needed 
is
$E[T_A]=N=n*k,$
because we would be testing each individual exactly once.

For Approach (B): 
\begin{enumerate}
\item
Let $Y$ = the \# of pools infected
and 
$T_B$ = the total number of tests needed.
Assuming independence, we have that
$E[Y]=n(1-.95^k)$ and
$E[T_B]=n+kE[Y]=n+k(n(1-.95^k))$.  
With $N=5000$ people, this simplifies to:
$E[T_B]=5000(1/k+(1-.95^k))$.    
When $k=10$, $n=500$ and $P(\textrm{infected})=p=0.05$, $E[T_B] = 2506.3$, 
which closely matches the results from the simulation.

\item
We find the derivative of $E[T_B]$ with respect to $k$ and solve (using a symbolic
mathematics package such as Maple or
Wolfram Alpha), which 
yields
a positive solution of $k=5.022$ (see Figure \ref{fig:plot1b}).  
When $k=5$, $$E[T_B]=n+5(n(1-.95^5)) =n+1.13n =2.13n = 2130.$$  
This result is
similar to that shown in 
the empirical simulations.

\item We compare the two expected number of tests needed for each of the approaches when $k=5$ and $p=0.05$:
\begin{eqnarray*}
E[T_A] / E[T_B] = 5n/2.13n = 2.35.
\end{eqnarray*}
Approach (A) requires an average of 2.35 times the number of tests than Approach (B).
Figure \ref{fig:plot1b} demonstrates that this ratio is greater than 1 for pool sizes between 2 and 10,
given a prevalence of $0.05$.
\end{enumerate}

\subsubsection{Commentary}
This problem was part of a series of probability questions 
at the start of the course.  While more efficient programming
approaches could be used, the empirical solution features a number of idioms and 
tricks of the trade that are repeated throughout the class.

This example demonstrates a setting where 
the analytic solution is straightforward using 
basic properties of expectations, but where the empirical solution provides
a useful check on the results.
This type of question helps students build confidence in using knowledge
from the prerequisite course in new ways.

\subsection{Sampling from a probability distribution} \label{pathological}

Questions [from \citeasnoun{evan:rose:2004}]:
\begin{enumerate}
\item Is it possible to find a tractable expression for the cdf of a distribution with 
density given by: $f(y) = c
(1+|y|)^3
\exp(-y^4)
,$ where 
$c$ is a normalizing constant and $y$ is defined on the whole real line?
If not, can you find $c$?
\item  Show how to generate a sample of observations from this distribution.
\item Describe how this is useful in Bayesian inference.
\end{enumerate}

\subsubsection{Solution}

\begin{figure}[bhtp]
\begin{center}
\includegraphics[height=5in,width=5in]{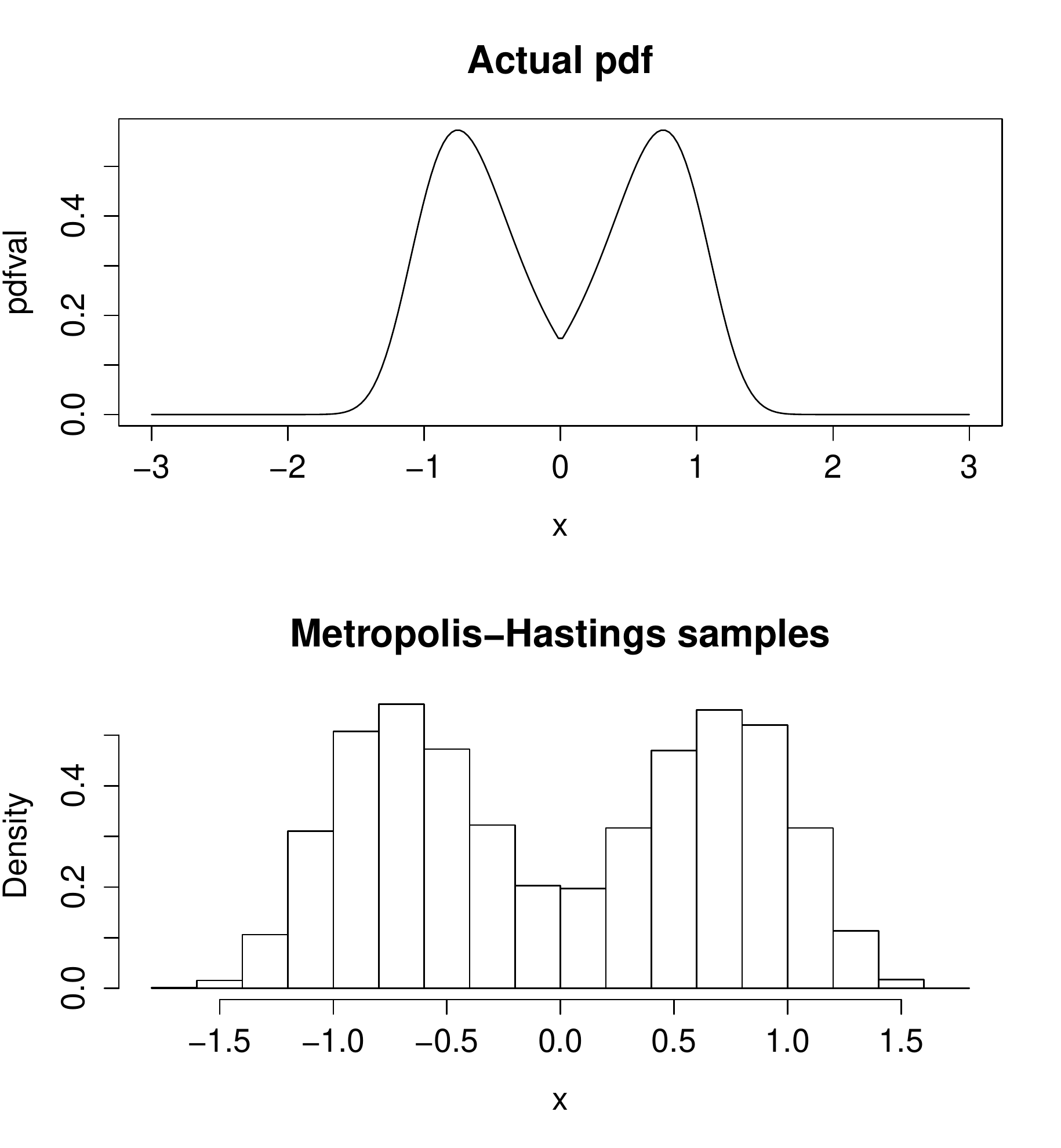}
\caption{True density and simulated draws from probability distribution}
\label{fig:samp1b}
\end{center}
\end{figure}

\begin{enumerate}
\item
While it is possible to find a closed-form solution for 
the cdf of this distribution it is not easily 
solvable. Note that because the density is a function of the
absolute value of $y$, the integral can be broken into two symmetric
parts.
To find $c$, we evaluate twice the integral from $[0, \infty)$
in R:
\begin{verbatim}
> f = function(x){exp(-x^4)*(1+abs(x))^3}
> integral = integrate(f, 0, Inf)
> 2 * integral$value
[1] 6.809611
\end{verbatim}
Hence $c=1/6.809611\cong 0.15.$

\item
We created a Markov Chain Monte Carlo sampler, using the Metropolis-Hastings algorithm. The premise for this algorithm is that it chooses proposal 
probabilities so that after the process has converged draws are generated
from the desired distribution. A further discussion for enthusiasts can be found on Page 610 of \citeasnoun{evan:rose:2004}.
We find the acceptance probability $\alpha(x, y)$ in terms of two densities, our $f(y)$ and $q(x,y)$, a normal proposal 
density with mean $x$ and variance 1, so that 
\begin{eqnarray*}
\alpha(x,y) &=& \mathrm{min}\left\{1, \frac{f(y) q(y, x)}{f(x) q(x,y)} \right\} \\
&=& \mathrm{min}\left\{1, \frac{c \exp{(-y^4)}(1+|y|)^3 (2 \pi)^{-1/2} \exp{(-(y-x)^2/2)}}{c \exp{(-x^4)}(1+|x|)^3 (2 \pi)^{-1/2} \exp{(-(x-y)^2/2)}}   \right\} \\
&=& \mathrm{min}\left\{1, \frac{ \exp{(-y^4+x^4)}(1+|y|)^3 }{(1+|x|)^3}    \right\} \\
\end{eqnarray*}

Pick an arbitrary value for $X_1$. The Metropolis-Hastings algorithm then computes the value $X_{n+1}$
as follows: \\
1. Generate $Y_{n+1}$ from a normal($X_n$, 1). \\
2. Let $y = Y_{n+1}$, compute $\alpha(x, y)$ as before. \\
3. With probability $\alpha(x, y)$, let $X_{n+1} = Y_{n+1} = y$ (use proposal value). Otherwise, with 
probability $1-\alpha(x, y)$, let $X_{n+1} = X_n = x$ (keep previous value). \\

The code (displayed in Figure \ref{fig:samp1a}), uses the first 100,000 iterations as a burn-in period, then generates 100,000
samples.  A histogram is displayed in Figure \ref{fig:samp1b}.

\item 
The Metropolis-Hastings algorithm is a form of Markov Chain Monte Carlo 
(MCMC) and is 
particularly attractive when the posterior density function 
does not have a familiar integral (such as when $f(x)$ is a posterior density that does not correspond to a conjugate prior).

Simulation is a central part of applied 
Bayesian analysis, because of the relative ease with which samples can be generated 
from a probability distribution, even when the density function cannot be explicitly integrated (see page 25 of \citeasnoun{gelm:carl:2004}).
\end{enumerate}

\begin{figure}[htpb]
\begin{verbatim}
x = seq(from=-3, to=3, length=200) 
pdfval = 1/6.809610784*exp(-x^4)*(1+abs(x))^3 
par(mfrow=c(2, 1)); plot(x, pdfval, type="n"); lines(x, pdfval)
title("Actual pdf")

alphafun = function(x, y) { return(exp(-y^4+x^4)*(1+abs(y))^3*
   (1+abs(x))^-3) } 
numvals = 100000; burnin = 100000 
i = 1 
xn = 3 # arbitrary value to start
while (i <= burnin) { 
   propy = rnorm(1, mean=xn, sd=1) 
   alpha = min(1, alphafun(xn, propy)) 
   unif = runif(1) < alpha 
   xn = unif*propy + (1-unif)*xn 
   i = i + 1 
} 
i = 1 
res = rep(0, numvals) 
while (i <= numvals) { 
   propy = rnorm(1, mean=xn, sd=1) 
   alpha = min(1, alphafun(xn, propy)) 
   unif = runif(1) < alpha 
   xn = unif*propy + (1-unif)*xn 
   res[i] = xn 
   i = i + 1 
} 
hist(res, nclass=20, main="Metropolis-Hastings samples", 
   xlab="x", freq=FALSE)
\end{verbatim}
\caption{R code to generate Metropolis-Hastings samples}
\label{fig:samp1a}
\end{figure}

\subsubsection{Commentary}

This problem was taken from the final set of problems, entitled \emph{Bayesian statistics: a big idea}, which
was intended to introduce students to more sophisticated simulations that are necessary to get answers
for more complex models.  
Because the students had no prior experience with MCMC, a preliminary
mini-lecture on the topic was provided along with supporting readings from \citeasnoun{lavi:2007}.  This included some classic examples with conjugate priors.
Throwing the nasty density function at students was initially off-putting, 
but it helped to motivate MCMC methods and introduce Bayesian ideas and methods.
The goal of this section was to
give students a glimpse into a flexible and sophisticated set of models that
can tackle problems far outside the realm of a traditional math stat class.

\section{Grading and assessment}
\label{sec:assessment}

Assessment of students in the course was done in several ways.
Students completed 7 sets of problems over the course of the semester
(each one approximately 2 weeks
apart).  
Grades on
preliminary solutions and weekly papers constituted 35\% of the
grade, with class participation, attendance and oral presentations
an additional 20\%.  Two midterm exams accounted for 40\% while 5\%
reflected good faith effort towards completion of low-stakes online assessments.
The midterm exams had in-class and take-home components.  
They included problems similar to those undertaken by the groups, albeit with simpler solutions.

An informal mid-semester evaluation was undertaken approximately 
halfway through the course.  For the first offering of 
the course, a colleague 
met with the class during the
last 15 minutes of a class session (without the instructor present).
Feedback from this assessment indicated great worries about the structure
of the take-home midterm (would the problems be as hard as Rice?)
and queries about other forms of assessment.

For the second offering, a more formal evaluation was undertaken where 
a staff member from the college learning center spent the last 20
minutes of a class session with students in focus groups.  The students
appreciated the structure of the course and the opportunities for 
revision.  They ``like that we get lectures on background, the collaboration
and group work'' and ``like that we do analytic and empirical solutions."
Students
sought more input from the instructor, with a desire for more 
lectures to ``put things into perspective."  Some students suggested
that the instructor ``tell us what are the key points to absolutely know
from each problem set."
The final question from the focus groups related to the students' roles
as learners.  Students revealed that they understand that they have to 
prepare more thoroughly for class, improve their own class participation
and assume additional responsibilities outside of class.  The students
acknowledged that they should read the text more carefully, read other 
groups' problems before the presentations, and ``try harder'' with Rice.

The outside evaluator summarized the report with the following quote:
\begin{quote}
As you made clear to me in our discussion, although your students may want you to tell them ``the key points to absolutely know,'' you believe strongly that they must work their way towards knowledge mastery in this course. To assist them in achieving this end, you have structured the course in ways that require them to work individually and collaboratively--with guidance from you--as they become more expert and reflective learners. 

Many of your students are uneasy with this approach and unsure of themselves: they want to know the \emph{right} answers, the \emph{correct} way to think, hence their request for more input from you. Their unease marks them as less sophisticated about real learning and/or timid about undertaking independent intellectual journeys. You might have an explicit discussion with your students about your pedagogy and your learning goals for them. I suspect they would be quite responsive to this kind of communication given their high regard for you and this course: they know you believe in them. And, since their answers to the third question reveal that they are aware of their own responsibilities
as students, you could also use this discussion to reinforce their own good insights on becoming more active and inquisitive learners.
\end{quote}

\begin{figure}
\begin{center}
\includegraphics[height=5in,width=5in]{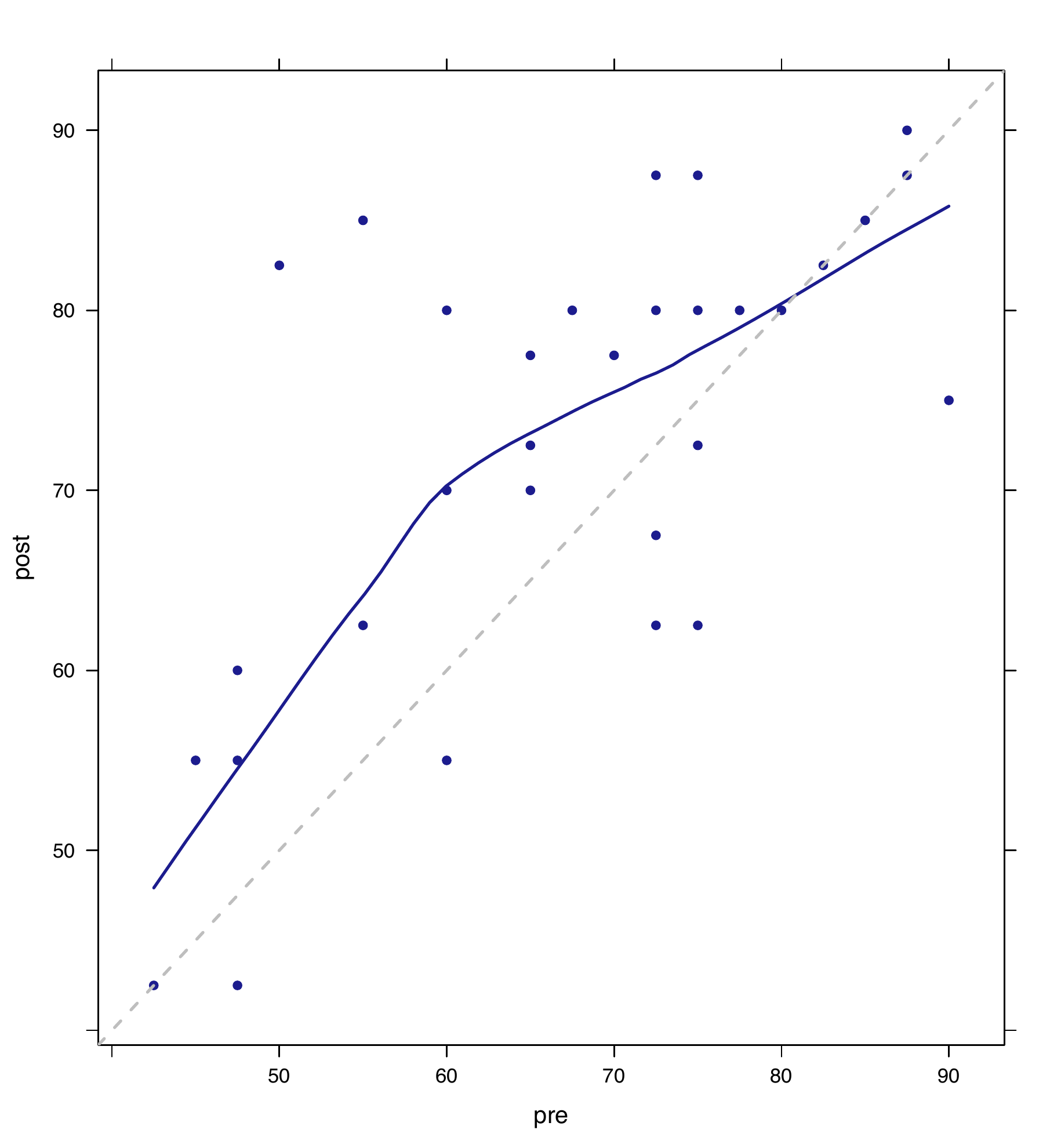}
\caption{
Relationship between 
student outcomes 
on the CAOS (Comprehensive Assessment 
of Outcomes in a first Statistics course) from
the class in 2007 and 2011
(plus smoothed line and $POST=PRE$ line).}
\label{fig:caos}
\end{center}
\end{figure}

The students also completed the CAOS post-test at the end
of the class, with a mean of 72.5\% correct (sd=13\%, min=43\%, max=90\%).  There was a statistically significant increase in
scores compared to the pre-test (paired t-test p=0.01, df=30, 95\% confidence interval from 1.4 to 9.2 point increase).  Figure \ref{fig:caos} displays
the relationship between pre and post scores (with a solid scatterplot 
smoother plus dashed $POST=PRE$ line).  There is some indication
of larger improvement for students with lower pre-test scores, which 
is consistent with a ceiling effect.  Given that the CAOS test is 
intended to assess outcomes from a first course, this is not surprising.

\section{Discussion}
\label{sec:discussion}

This paper describes an implementation of a modified Moore-Cohen method
mathematical statistics course at an undergraduate liberal arts college.
The course featured a series of challenging problems, some theoretical, others
data-driven, designed to help teach mathematical statistics using applications.  
A key idea is that 
the use of technology (R/RStudio and reproducible analysis tools)
has opened up new possibilities.

An attractive aspect of the proposed course was how it intentionally dovetailed with 
the GAISE recommendations \cite{gaise}.  In particular, it was designed to
encourage  statistical thinking 
through empirical problem solving, use real data to motivate methods,
stress conceptual understanding, foster active learning and use
technology to develop conceptual understanding.
The course is consistent with the American Statistical Association guidelines for
statistics programs \cite{asa-undergrad}, which call for students to develop effective
technical writing, presentation skills, teamwork and collaboration, in addition to 
knowledge of statistics.

\subsection{Comparisons, advantages and limitations}

This approach differs significantly from the traditional Moore-method, which 
was developed for a definition-theorem-proof type course and relies
primarily on
individual work and competition as a motivator.  
The modified Moore-Cohen method
uses group work to facilitate engagement, with stronger students able to plunge
more deeply into their solutions while still ensuring that weaker students
can receive assistance as needed.  
This modification might be better thought of as a species split-off, where 
rather than competing, 
students are supported to go beyond their expectation and discover something in
themselves.

A primary challenge of teaching is to engage students in the material
being studied.  \citeasnoun{cohe:1992} noted that \emph{the method
effectively raises the level of communication between students} and that
\emph{most students are stimulated by the change from passive to active 
learning}.  \citeasnoun{laza:2011} described the importance of capstone courses
in statistics.  Structuring the class with multiple, challenging 
problems that were not amenable to quick individual solution helped to
achieve the goals of a capstone.  This includes
getting students to grapple with real-world problems, helping them
develop capacities to work effectively in groups, augmenting their ability to 
compute to 
extend their problem-solving abilities, and helping them to sharpen their
abilities to communicate the complexity
and power of statistical methods.  The course also dovetails with
other efforts to involve students in interdisciplinary research projects 
\cite{legl:2010}, which tend to focus on larger, more complex datasets in the context
of a client discipline.

While no formal assessment of the course was undertaken, student
feedback from less formal appraisals  was generally positive.
Students found the 
 approach to be challenging, particularly 
at the beginning of the semester when they were confronted with 
simultaneously learning
R/RStudio, \LaTeX/Sweave, empirical problem solving techniques as well as
oral and written presentation skills.  
The particular technical challenges of learning new packages and
systems quickly receded, and the primary challenge related to answering
difficult questions and learning new material, concepts and statistical methods.

A limitation of problem or case-based courses is that they typically cover
fewer topics in more depth.  That was true for this course, which had
more constrained coverage goals than the traditional math stat class
(though most of the typical key concepts were covered).
In addition, students would be expected to have variable mastery of particular topics that were
included, since they engaged in the problems that their group was assigned
at an intense level, but had more passive involvement in the problems that
other groups presented.  The combination of written and oral 
presentation of solutions from other groups was designed to minimize these
disparities.  Ideally students would emerge from the course with useful
capacities (such as ability to compute with data, simulate to approximate
answers, and communicate orally and through their writing) that would allow
them to fill any gaps in their knowledge and succeed in a graduate level
course.

Classes that include group work 
as a component of assessments often have group dynamic issues, 
and this course was no
exception.  In general there was a positive sense of community and
engagement which flowed from the group-based workload.  
Knowing that the 
groups would be reshuffled twice helped as well.
Focusing much of the work in groups
allowed students to tackle far more challenging questions than
they could solve individually and also modeled a common
post-college work environment.  Several students have provided anecdotal
reports of the value of learning tools for statistical computing and reproducible analysis.

There are other challenges to use of this method for teaching the
mathematical statistics courses.  The enrollments were 12 and 20 students in  Spring 2007 
and Spring 2011, respectively.
While scaling to course sizes of 30--40 students would
be straightforward, larger class sizes would require different
systems and structures.  These might include multiple sections
taught with some common mini-lectures, doubling up on problems or
student support for computing.
 The time commitment was comparable to a standard
course, due to the extensive coaching and preparation, despite the fact that
formal lectures were relatively short (generally at the
start of each new topic).

\subsection{Use of technology}

Empirical (simulation-based) estimation complements analytic solutions,
and can often allow approximate solution of extremely challenging problems.  
Besides providing a useful check on analytic answers, these simulations
can help with insights into how to solve a problem.  R and RStudio serve as a flexible
and powerful environment for such exploration.  

A number of technologies were prominently featured in the course.
These included extensive use of \LaTeX\ and R. Reproducible analysis
(the {\tt Sweave} system \cite{leis:2002} as implemented within RStudio)
greatly facilitated integration of commands, output and graphics, and led to better
facility for students to undertake analyses outside the course.  This scaffolding
also helped to move students from a ``point and click'' approach to statistical 
analysis towards a more flexible scripting interface.  Further
discussion of how to integrate reproducible analysis and effective mechanisms
to build students' ability to ``compute with data'' are important issues but 
lie somewhat outside the scope of this paper.

Other courses may find the use of R and RStudio for simulation and approximation of
analytic solutions to be helpful, without the Moore-method approach.
The new text by \citeasnoun{prui:2011} features such a presentation.

\subsection{Closing thoughts}

\citeasnoun{cobb:2011} argues that the profession needs two types of 
statisticians: those with the capacity to appropriately analyze and interpret data, 
as well as those with interest in devising novel solutions to methodological challenges.
Teaching mathematical statistics in this manner has the potential
to foster engagement by presenting students with extended glimpses
of the excitement of developing statistical procedures to solve
challenging problems \cite{nola:temp:2010}.  This approach could
also serve as a model for other intermediate and advanced undergraduate
statistics classes.
This method may also be relevant for the teaching
of similar quantitative courses in other disciplines.

\section*{Acknowledgements}

This work was supported by NSF grant 0920350 (Phase II: Building a Community around Modeling, Statistics, Computation, and Calculus).  
Thanks to
Sarah Anoke,
George Cobb,
David Cohen, 
Daniel Kaplan,
David Palmer and 
Randall Pruim  for many useful discussions about pedagogy as well as
helpful comments 
on an earlier draft.  I am also indebted to the Editor, Associate
Editor and anonymous reviewers for many suggestions which led to improvements
in the manuscript.

\newpage
\singlespacing
\bibliographystyle{dcu}
\bibliography{hthesis}

\newpage

\section*{Online Appendix: Additional Example Problems}

I hear, I forget.  I do, I understand: \\ 
a modified Moore-method mathematical statistics course

The following material is proposed as an online appendix.

\subsection{Estimating $\sigma$ using IQR} \label{iqr}

Assume that we observe $n$ iid observations from a normal distribution. 
Questions:
\begin{enumerate}
\item Use the IQR of the list to estimate $\sigma$.
\item Use simulation to assess the variability of this estimator for samples of $n=100$ and $400$.
\item How does the variability of this estimator compare to $\hat{s}$ (usual 
estimator)?
\end{enumerate}

\begin{figure}[htpb]
{\small
\begin{verbatim}
numsim=1000; mu=42; n1=100; n2=400
runsim = function(numsim, n, mu, sigma) {
  res1 = numeric(numsim); res2 = res1
  for (i in 1:numsim) {
    mynorms = rnorm(n, 0, sigma)
    vals = quantile(mynorms)
    res1[i] = (vals[4]-vals[2])/1.34898
    res2[i] = sd(mynorms)
  }
  return(data.frame(IQR=res1, S=res2))
}
res100 = runsim(numsim, n1, mu, pi)
res400 = runsim(numsim, n2, mu, pi)
boxplot(res100$IQR, res100$S, res400$IQR, res400$S,
  names=c("n=100 (IQR)","n=100 (S)",
  "n=400 (IQR)", "n=400 (S)"),
  ylab="distribution of sigma-hat")
text(3.5, 4.0, "True sigma is 3.14159")
abline(h=pi); abline(v=2.5)
\end{verbatim}
}
\caption{R code to carry out simulation study (estimation of $\sigma$)}
\label{fig:tildecodea}
\end{figure}
\begin{figure}
\begin{center}
\includegraphics[height=5in,width=5in]{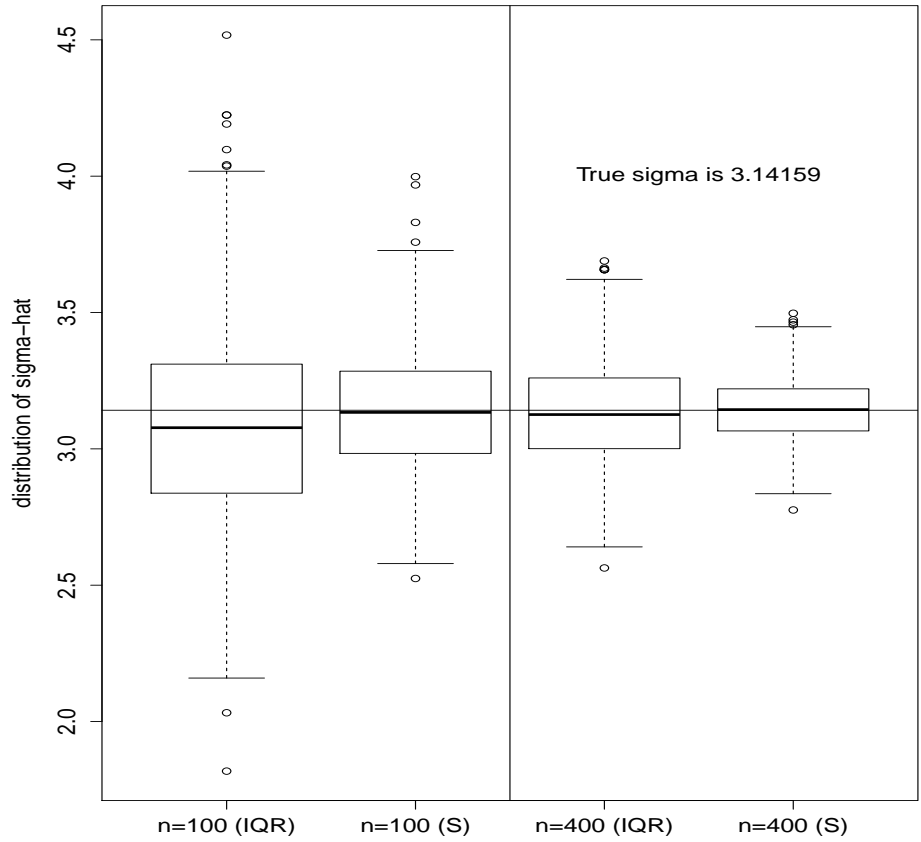}
\caption{Distribution of sample estimates by estimator and sample size}
\label{fig:tildeplotb}
\end{center}
\end{figure}

\subsubsection{Solution}

\begin{enumerate}
\item
We know that for a standard normal random variable $P(Z>0.675)=0.25$.  So we would 
expect that the IQR (interquartile range) would extend to $2*.6745=1.35$ standard units.  We use this 
expectation to determine the estimator: $\tilde{s}=IQR/1.35$.

\item
We carried out a simple simulation study with a fixed mean and standard deviation
(set to $\pi$).  A thousand simulations of samples were taken using $\tilde{s}$ and
$\hat{s}$ (MLE).  The results are displayed in Figure \ref{fig:tildeplotb}.
We note that both estimators are less variable when $n=400$ than for
$n=100$ and conclude that the variability of the estimators
goes down as a function of $\sqrt{n}$.

\item 
The IQR for $\tilde{s}$ is 0.50 for n=100 and 0.25 for n=400, while the IQR for $\hat{s}$ is 0.30 for n=100 and 0.16 for n=400.  
We conclude that the MLE is more efficient than our ad-hoc estimator.

\end{enumerate}

%

\subsubsection{Commentary}

This exercise was included with a problem set mid-way through the class
as the
nature and properties of estimators were explored.
This problem introduced the idea of a simulation study to investigate
the behavior of a new estimator.  While the analytic solution was
straightforward, it required the students to think about estimation
in a different way, and tap properties of the normal distribution.  The empirical solution provided a glimpse into
the additional variability of the IQR estimator compared to
the standard estimator of standard deviation.  A full analytic solution for this problem
was beyond the scope of the course, but can be undertaken for specific values of $n$.

\subsection{Assessing robustness of chi-square statistic to small cell counts} \label{chisquare}

Perform a simulation study on the sensitivity of the $\chi^2$ test for the uniform distribution to expected cell counts below 5. Simulate the distribution of the test statistic for 16 and 64 observations from a uniform distribution using 8 equal-length bins (from \citeasnoun{nola:spee:2000}).

\begin{figure}[htpb]
{\small
\begin{verbatim}
simchisq = function(n, bins) { 
  vals = cut(runif(n, 0, bins), breaks=0:bins)
  obs = c(table(vals))
  exp = c(rep(n/bins, bins))
  return(sum(((obs - exp)^2)/exp))
}
library(mosaic); par(mfrow=c(1, 2))
bins = 8; n = 16    # Expected count per cell equal to 2
res = do(10000)*simchisq(n, bins)
plot(density(res$result), main="", lwd=2, 
     xlab=paste("N=",n,", ",bins," bins", sep=""), xlim=c(0, 20))
curve(dchisq(x, bins-1), 0, max(res$result), add=TRUE, lwd=2, lty=2)

n = 64              # Expected count per cell equal to 8
res = do(10000)*simchisq(n, bins)
plot(density(res$result), main="", lwd=2, 
  xlab=paste("N=",n,", ",bins," bins", sep=""), xlim=c(0, 20))
curve(dchisq(x, bins-1), 0, max(res$result), add=TRUE, lwd=2, lty=2)
\end{verbatim}
}
\caption{R code to carry out simulation study (chi-square problem)}
\label{fig:tildecode2}
\end{figure}
\begin{figure}
\begin{center}
\includegraphics[height=5in,width=5in]{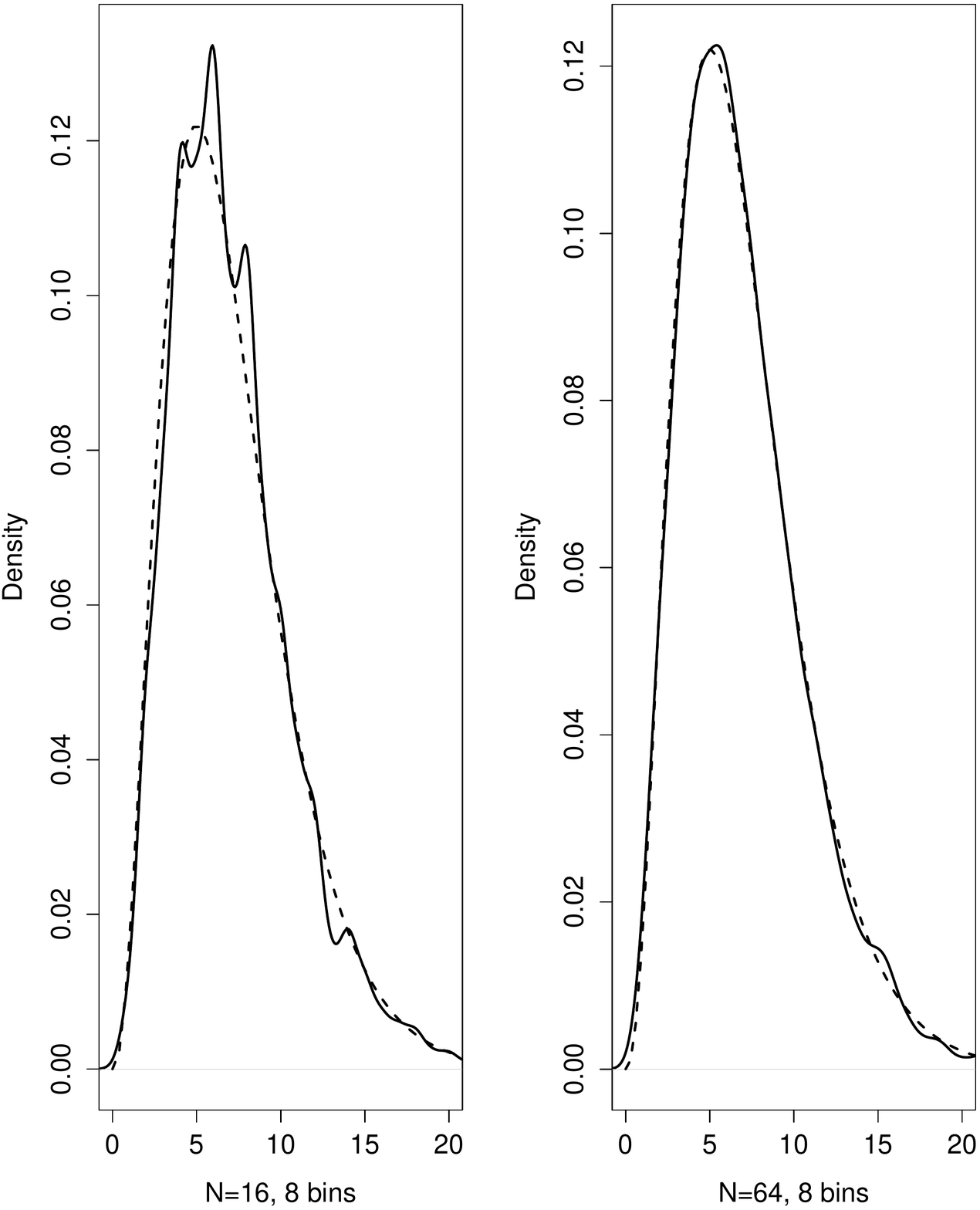}
\caption{Observed and expected distribution for chi-square statistic}
\label{fig:tildeplot2}
\end{center}
\end{figure}

\subsubsection{Solution}

We know that the chi-square test is recommended only in situations where the expected cell
count is 5 or more in each cell.  In this simulation study, we generate repeated samples 
from the null distribution and compare these to the large-sample distribution of the
chi-square ($\chi^2$) statistic (see Figure \ref{fig:tildecode2}).  
We know that in this setting, the appropriate degrees of freedom are equal to the number of
bins minus 1.  The main work is done using the {\tt simchisq()} function, which generates
data from a continuous uniform variable, then constructs the observed and expected cell
counts and the chi-square statistic.  This is repeated for the two scenarios and 
displayed in 
Figure \ref{fig:tildeplot2}.
We see that the observed distribution under the null is somewhat jumpy (due to the
discreteness of the possible values) when the expected cell counts are low (left figure), and that
the observed curve is quite similar to the chi-square distribution when the expected
cell count is 8 (right figure).

\subsubsection{Commentary}

This problem was intended to provide more practice in the construction of 
simulation studies as well as introduce new idioms related to looping and 
writing of functions.  It also serves to highlight the importance of assumptions
and the idea of sampling under the null distribution (as a precursor to resampling
based inference).
This was included with a group of problems mid-way through the class
as the
nature and properties of tests of hypotheses along with sampling distributions
under the null were explored.

\end{document}